\documentclass{ws-procs975x65}
\usepackage{graphicx}

\def\beq{\begin{equation}}
\def\eeq{\end{equation}}

\begin{document}

\title{Pulse-wise GRB correlation: implication as a cosmological tool}
\author{Rupal Basak$^*$}

\address{Nicolaus Copernicus Astronomical Center,\\
Bartycka 18, PL-00-716 Warszawa, Poland\\
$^*$E-mail: rupal.basak@gmail.com}

\author{A. R. Rao} 

\address{Tata Institute of Fundamental Research,\\
Homi Bhabha Road, Mumbai-400005, India\\
E-mail: arrao@tifr.res.in}

\begin{abstract}
Gamma-ray bursts (GRBs) are cosmological explosions which carry valuable information 
from the distant past of the expanding universe. One of the greatest discoveries in
modern cosmology is the finding of the accelerated expansion of the universe using 
Type Ia supernovae (SN Ia) as standard candles. However, due to the interstellar
extinction SN Ia can be seen only up to a redshift $z\sim 1.5$. GRBs are considered 
as the potential alternative to push this limit to as high as $z\sim 10$, a redshift
regime corresponding to an epoch when the universe just started to form the first structures. There exist several 
correlations between the energy and an observable of a GRB which can be used to derive
luminosity distance. In recent works, we have studied spectral evolution within the individual 
pulses and obtained such correlations within the pulses. Here we summarize our results of the pulse-wise GRB
correlation study. It is worth mentioning that all GRB correlations are still empirical, 
and we cannot use them in cosmology unless we understand the basic physics of GRBs.
To this end, we need to investigate the prompt emission spectrum which is so far 
generally described by the empirical Band function. We shall discuss our current understanding 
of the radiation process particularly the finding of two blackbodies and a powerlaw 
(the 2BBPL model) as the generic spectral model and its implication. This is a work in 
progress and we expect to obtain the most fundamental GRB correlation based on our 
improved spectral model.\end{abstract}


\bodymatter


\section{Introduction}
Gamma-ray bursts (GRB) are cosmological explosions which produce an enormous amount 
of energy $\sim 10^{50}-10^{53}$\,erg within a few seconds to a few minutes by means 
of a highly relativistic jet.
A GRB light curve generally consists of a single or multiple broad pulses and rapid
variability within the pulses on a timescale down to a few milliseconds.\cite{Piran2004}
On the other hand, the spectrum is a continuum which is generally attributed 
to an optically thin synchrotron emission.\cite{Piran2004,Rees1992} Interestingly, the spectrum generally 
shows a hard to soft evolution within the individual pulses.\cite{Hakkila2008}
Hence, pulses are important and should be considered independently.

\begin{figure}\centering
\begin{center}
\includegraphics[width=\textwidth]{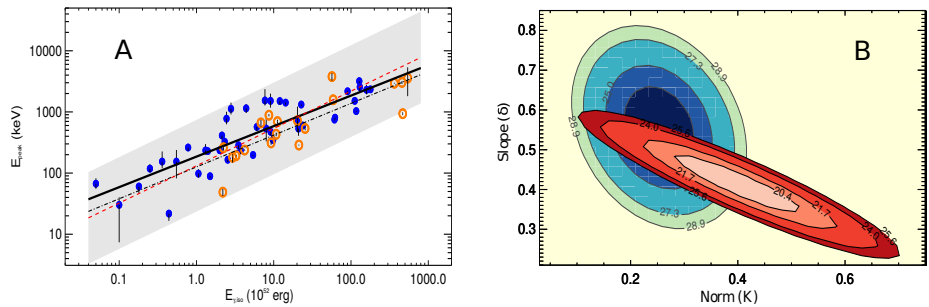}
\end{center}
\caption{(A) $E_{\rm peak}$--$E_{\rm iso}$ correlation: Pulse-wise (blue filled circles fitted with
black solid line), burst average correlation of the same GRBs (orange open circles fitted with 
violet dot-dashed line) and the same from Nava catalog \cite{Nava_2012} (red dashed line with shaded $3\sigma$ data scatter).
(B) The $\chi^2$ contours of the fit parameters of the pulse-wise correlation for two redshift bins. \cite{rr2013_correlation}}
\label{GRB_correlation}
\end{figure}

In recent works, \cite{rr2012_090618, rr2014_singlepulse} we 
have given particular attention to the individual GRB pulses and obtained the 
timing and spectral informations. We have then used the GRB pulses to study 
GRB correlations\cite{rr2012_correlation, rr2013_correlation} similar to the
Amati correlation\cite{Amati_2002}. Such studies are 
important due to several reasons: (1) As the correlation hold within 
the pulses it shows that the correlation is a pulse property, (2) it also 
shows that the correlation is real and not due to selection bias, (3) it
provides a larger sample compared to the burst average studies.
Though pulse-wise GRB correlation is powerful and indicates a basic mechanism 
of pulse generation, our pulse description is still empirical. Here we review 
the results of our correlation studies and the possible
improvements based on our new findings on the GRB spectrum.

\section{Pulse-wise GRB correlation}
We have developed a pulse model which describes a pulse simultaneously in energy 
and time domain.\cite{rr2012_090618} For this we use the spectral model as Band 
function\cite{Band1993}, lightcurve as Norris model\cite{Norris2005} and 
spectral evolution as Liang-Kargatis model\cite{lk96} (hereafter LK96). Our model 
is implemented in the X-ray spectral fitting package {\tt XSPEC} as a table model
with two parameters --- the peak energy at the beginning of a pulse ($E_{\rm peak,0}$)
and the characteristic evolution parameter ($\phi_0$). We apply this model on GRB 
090618 and generate the individual pulses. We derive various pulse 
properties e.g., lightcurve and pulse width in various energies, spectral 
lag. A comparison with the data shows an excellent agreement. We then use this model 
for a set of \emph{Fermi} GRBs with known redshift ($z$). We find that 
the $E_{\rm peak,0}$ bears a strong correlation with the energy ($E_{\rm \gamma, iso}$)
within the individual pulses with a Pearson coefficient, $r=0.96$ and a 
chance probability, $P=1.6\times10^{-12}$.

In a recent work,\cite{rr2013_correlation} we have studied pulse-wise Amati 
correlation for a set of 19 GRBs with 43 pulses. We again obtain a good correlation
(see Fig.~\ref{GRB_correlation}A). One of the major goal of GRB correlation 
study is their application to constrain the cosmological parameters like 
matter and dark energy density at high redshifts, $z\sim10$ compared to that 
provided by the type Ia supernovae ($z<1.5$ in optical wavelengths). However, 
we need to investigate whether the GRB correlation holds for different redshifts. 
In Fig.~\ref{GRB_correlation}B, we plot the contours ($1\sigma$ for single parameter
and $1-3\sigma$ for two parameters) of the correlation 
parameters for different $z$ bins, and we clearly see that they
match within $1\sigma$. Hence, our correlation 
is stable with $z$.

\begin{figure}\centering
\begin{center}
\includegraphics[width=\textwidth]{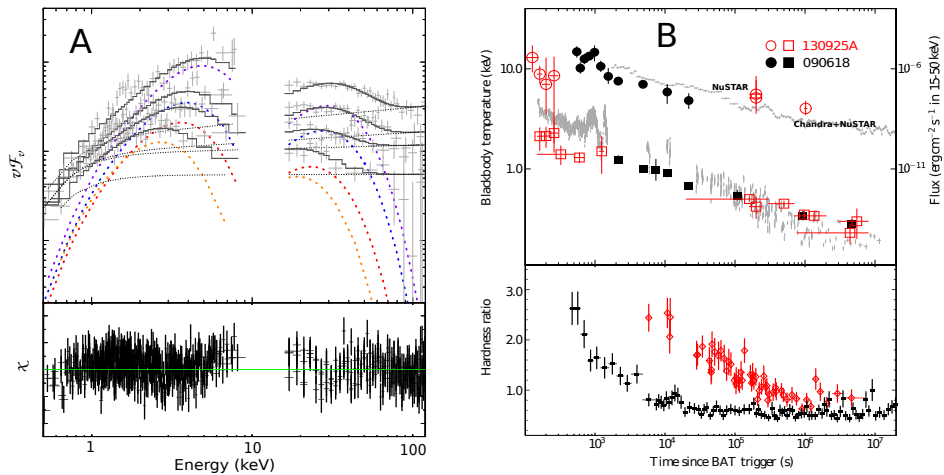}
\end{center}
\caption{(A) Time resolved BAT--XRT data of GRB 090618 fitted with the 2BBPL model (upper panel) and residual (lower panel).
The two blackbodies are shown by dotted lines with different colors for different time bins. 
(B) Evolution of the temperature of the 
two blackbodies (upper panel) and the hardness ratio (lower panel) for two GRBs. \cite{rr2015_090618, rr2015_130925A}}
\label{2BBPL}
\end{figure}

\section{Radiation Mechanism --- Thermal Component}
The pulse model we have developed is an empirical one. It is worthwhile to 
mention that all GRB correlations are empirical and shows considerable data 
scatter. This is also found in our pulse-wise correlation study (Fig.~\ref{GRB_correlation}A). Hence, it is 
important to understand the physical mechanism of pulse emission and spectral
characteristics. In some studies, the synchrotron origin of the spectrum has been 
challenged by observations of a spectral photon index higher than -1.5 (or even higher 
than -2/3), which is disallowed in the fast cooling (slow cooling) regime.\cite{Preece1998} 

In recent works it is found that the GRB spectrum also contains a thermal  
component.\cite{Ryde2004} In general the spectrum is a combination of the thermal 
and non thermal emission. However, the correct shape of the components is debatable,
e.g., a blackbody+powerlaw (BBPL)\cite{Ryde2004}, multi-colour blackbody+powerlaw
(mBBPL)\cite{Ryde2010}, two blackbodies+powerlaw 
(2BBPL)\cite{rr2014_singlepulse,rr2013_parametrized,rr2014_manypulse,rr2015_090618,rr2015_130925A},
2BB+cutoff powerlaw (2BBCPL)\cite{rr2015_090618} etc. Moreover, the thermal emission is 
not always statistically significant. One way to find the correct model is to analyze the 
data with good spectral resolution e.g., \emph{Swift} X-ray telescope (XRT), \emph{NuSTAR},
\emph{Chandra}. However, due to the focusing nature of these instruments such observations 
are rare. Recently, we have found two such cases, GRB 090618\cite{rr2015_090618}
and GRB 130925A.\cite{rr2015_130925A} In  Fig.~\ref{2BBPL}A, we have shown the 
2BBPL model fitted to the time resolved data of GRB 090618 obtained from \emph{Swift}
burst alert telescope (BAT) and the XRT. In Fig.~\ref{2BBPL}B the temperature and hardness
evolution of this GRB is compared with that found for GRB 130925A observed with 
the XRT, the \emph{NuSTAR} and the \emph{Chandra}. We see a remarkable similarity of the spectral 
evolution which denotes a common radiation mechanism.

\begin{figure}\centering
\begin{center}
\includegraphics[width=\textwidth]{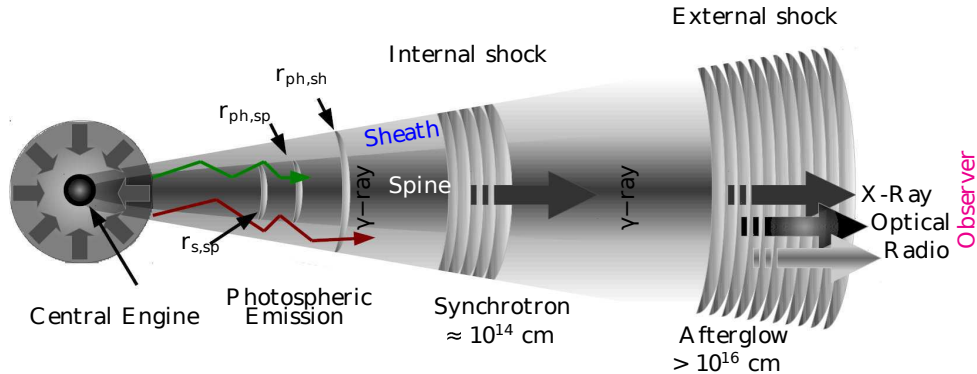}
\end{center}
\caption{A schematic of the spine-sheath jet \cite{rr2015_090618} (see text for detail).}
\label{jet}
\end{figure}

We note here that the pulse emission in GRBs can have a wide variety. We have studied 
all such varieties and found a clear evidence of the two blackbodies in all cases,
namely, (1) GRBs with single pulse\cite{rr2014_singlepulse}, (2) GRBs with separable 
multiple pulses\cite{rr2013_parametrized} and (3) those having highly variable
lightcurve\cite{rr2014_manypulse}. Hence, the model appears to be a generic spectral 
shape.

We suggest a spine-sheath jet structure to explain our observations
(see Fig.~\ref{jet}).\cite{Ramirez-Ruiz2002,Zhang2004}
Such a structure is theoretically expected as the GRB jet pierces through the
envelop of the progenitor star. The two blackbodies are produced at the
two photospheres of the jet, while the photons crossing the boundary layer
of the spine and sheath are inverse-Compotonized and form a cutoff powerlaw.
In addition, internal shocks produced at larger radius will also contribute 
to the non-thermal emission. We are planning to pin down the mechanism by
developing a spine-sheath jet model\cite{Ito2014} and then applying this 
model on a large set of GRBs to obtain the physical parameters of the jet.

\begin{figure}\centering
\begin{center}
\includegraphics[width=\textwidth]{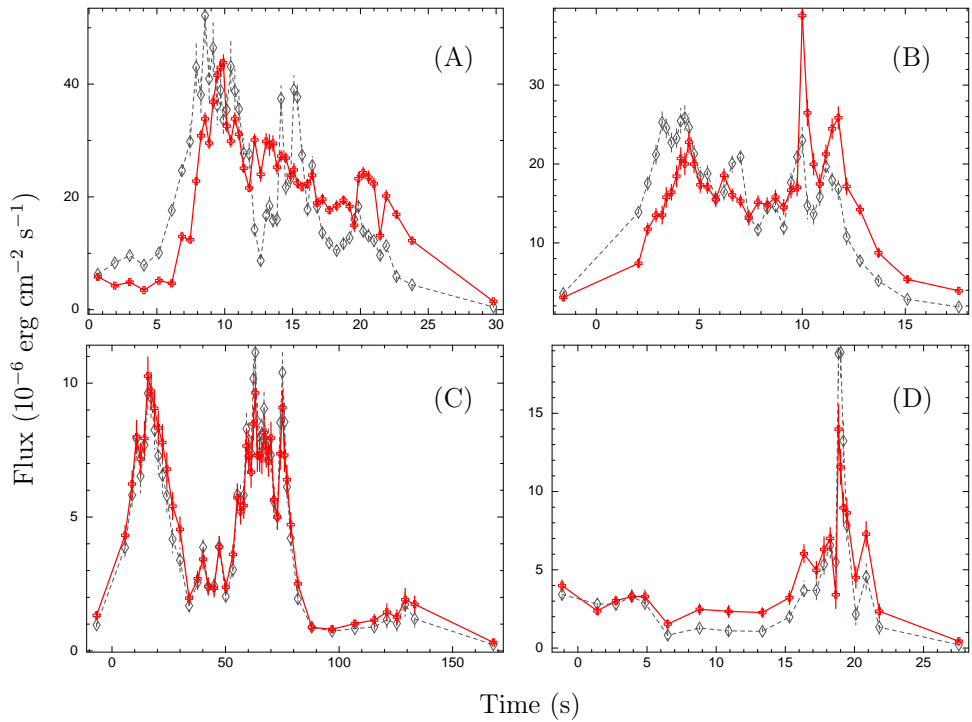}
\end{center}
\caption{Flux evolution of the total (black dashed lines) and the powerlaw (red solid lines) for GRBs with 
LAT detection.\cite{rr2013_GeV} GRBs with high LAT flux (A, B), and low LAT flux (C, D)}
\label{delayed_PL}
\end{figure}

\begin{figure}\centering
\begin{center}
\includegraphics[width=\textwidth]{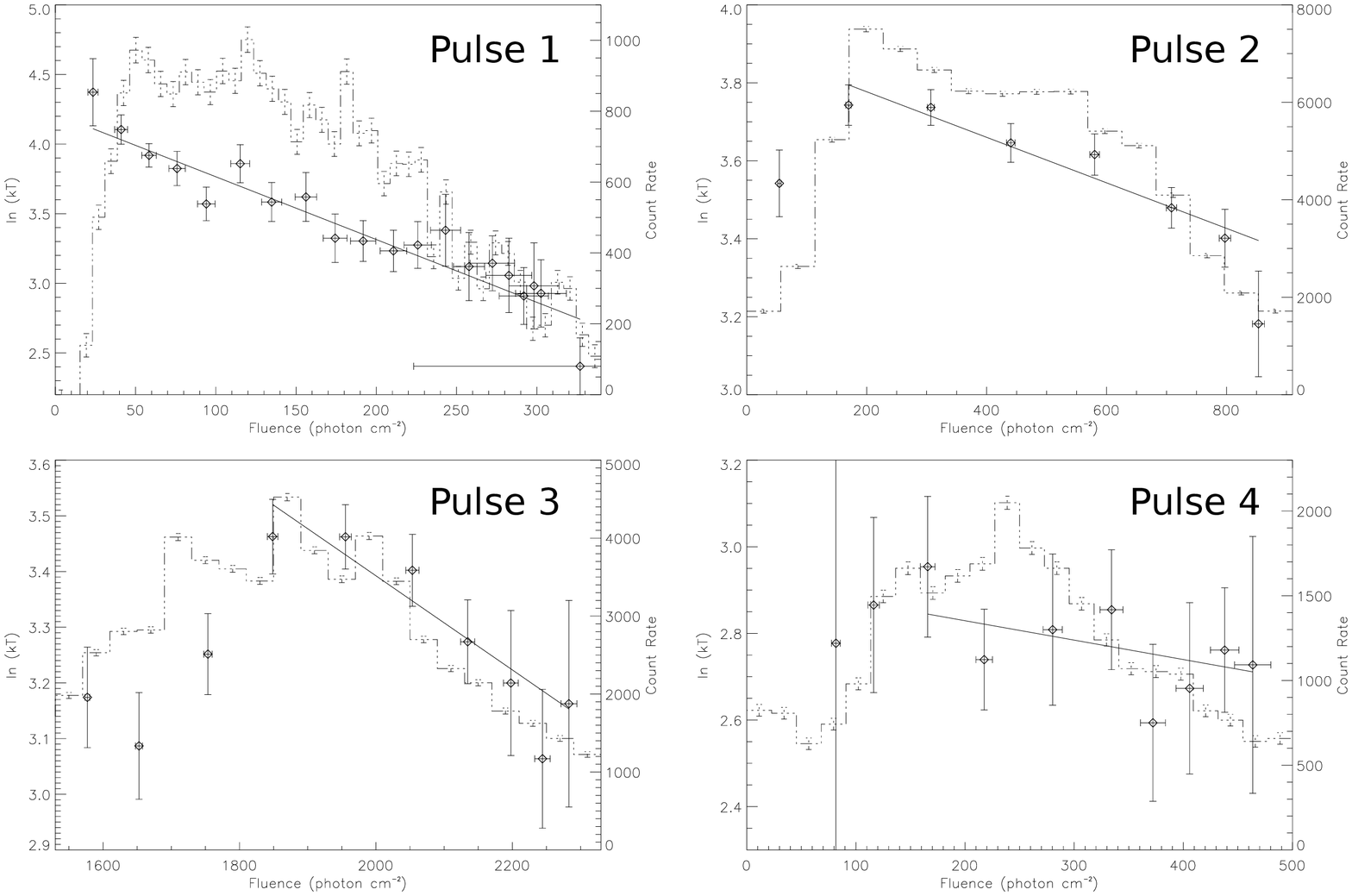}
\end{center}
\caption{Temperature evolution similar to the LK96 law \cite{lk96} for the pulses of GRB 090618. The evolution 
is shown by the fitted solid lines while the count rates are shown by dot-dashed line.}
\label{lk96}
\end{figure}

\section{Discussion}
The 2BBPL model is primarily found to be consistent with the data in keV--MeV 
or lower (e.g., the XRT) energies. It is useful to apply this model 
to the prompt emission data in these energies and then investigate the 
predicted flux in other energies. In the following we perform such analysis 
to see how much we can explain the emission at very high energies (GeV) 
based on the spectral model fitted to the lower energy data.

The very high energy (GeV) emission is detected in some GRBs with \emph{Fermi} 
large area telescope (LAT). This emission is delayed compared to the prompt  
keV--MeV emission by a few seconds. It is well known that the Band function cannot 
capture the wide spectrum from keV to GeV. Also, the GeV flux is found to be uncorrelated
with the prompt keV--MeV flux. It is interesting to fit the keV--MeV data with the 2BBPL model
and then compare with the flux at GeV energies. Analyzing a set GRBs detected by 
the LAT, we find a significant correlation of the GeV photon fluence with that of 
the non-thermal (powerlaw) component of the 2BBPL model fitted to 
the keV--MeV data.\cite{rr2013_GeV} Also, the powerlaw flux of GRBs with high
GeV emission tend to have a delayed onset, and this component lingers at the final phase of the
prompt emission (see Fig.~\ref{delayed_PL}). Remembering that the GeV emission is delayed and long lasting than the keV--MeV
emission, we strongly suggest that the powerlaw component of the 2BBPL model shares a common
origin with the GeV emission. In the framework of the spine-sheath jet model, this emission is produced 
at higher radius probably by synchrotron emission. On the other hand, the GeV photons of 
GRBs with low GeV emission are probably produced due to inverse-Compton of photons crossing 
the spine-sheath boundary. Clearly such finding strongly validates the 2BBPL model.

In future we shall test whether one or more spectral components of the 2BBPL model 
can be used to study the GRB correlation. For example, in our pulse model we have assumed 
that the spectral evolution is given by the LK96 law. We have found that fitting the 
spectrum with an additional blackbody component leads to a similar evolution. Now,
the blackbody temperature shows a similar behaviour as shown by the $E_{\rm peak}$
(see Fig.~\ref{lk96}). Such findings give us hope to further improve the pulse 
description and finally obtain the fundamental GRB correlation based on the physical 
spectral model.

\section*{Acknowledgement}
RB is a stipendiary of START program of the Polish Science Foundation (2016)
and supported by Polish NCN grants 2013/08/A/ST9/00795, 2012/04/M/ST9/00780,
2013/10/M/ST9/00729, and 2015/18/A/ST9/00746.

\end{document}